\documentclass[showpacs,aps]{revtex4}
\usepackage{amssymb}
\usepackage{amsmath}
\usepackage{graphicx}
\usepackage{lscape}
\usepackage{booktabs}
\usepackage{subfig}

\begin{document}

\title{Systematic study of elliptic flow parameter in the relativistic
        \\ nuclear collisions at RHIC and LHC energies}
\author{Ben-Hao Sa$^1$ \footnote{sabh@ciae.ac.cn}, Dai-Mei Zhou$^2$,
        Yu-Liang Yan$^1$, Yun Cheng$^2$, Bao-Guo Dong$^1$, Xu Cai$^2$}

\affiliation{$^1$ China Institute of Atomic Energy, P. O. Box 275 (10),
              Beijing, 102413 China. \\
             $^2$ Key Laboratory of Quark and Lepton Physics (MOE) and
              Institute of Particle Physics, Central China Normal University,
              Wuhan 430079, China.}

\begin{abstract}
We employed the new issue of a parton and hadron cascade model PACIAE 2.1 to
systematically investigate the charged particle elliptic flow parameter $v_2$
in the relativistic nuclear collisions at RHIC and LHC energies. With randomly
sampling the transverse momentum $x$ and $y$ components of the particles
generated in string fragmentation on the circumference of an ellipse instead
of circle originally, the calculated charged particle $v_2(\eta)$ and
$v_2(p_T)$ fairly reproduce the corresponding experimental data in the
Au+Au/Pb+Pb collisions at $\sqrt{s_{NN}}$=0.2/2.76 TeV. In addition, the
charged particle $v_2(\eta)$ and $v_2(p_T)$ in the p+p collisions at $\sqrt
s$=7 TeV as well as in the p+Au/p+Pb collisions at $\sqrt{s_{NN}}$=0.2/5.02
TeV are predicted.
\end{abstract}
\pacs{25.75.-q, 24.10.Lx}
\maketitle
%%%%%%%%%%%%%%%%%%%%%%%%
\section {Introduction}
%%%%%%%%%%%%%%%%%%%%%%%%
To explore the phase transition from the hadronic matter (HM) to quark-gluon
matter (QGM) is one of the fundamental aims of relativistic nuclear
collisions. A couple years ago, four international collaborations of BRAHMS,
PHOBOS, STAR, and PHENIX at RHIC have published white papers
\cite{brah,phob,star,phen} to declare their evidences for the discovery of
strongly coupled quark-gluon plasma (sQGP). One of the most important signals
is the large elliptic flow parameter of produced particle in the Au+Au
collisions at $\sqrt{s_{NN}}$=200 GeV.

The measurement of particle elliptic flow parameter $v_2$ is not trivial.
Several methods have been proposed, such as the event plane method \cite{posk},
Lee-Yang zero point method \cite{lyzp}, and the cumulant method \cite{cumu}
etc. The cumulant method is even distinguished with two-, four-, and
six-particle cumulants. The discrepancy among the $v_2$ values measured with
the event plane method, Lee-Yang zero point method, and the cumulant method
may reach a few ten percent as shown in Fig. 4 and 5 of \cite{star1} and in
Fig. 11 of \cite{cms}. Recently, one even argued that the event plane method
is obsolete \cite{luzu}.

On the other hand, the particle elliptic flow parameter $v_2$ is also not
easy to investigate theoretically. The conventional (hadronic) transport
(cascade) models always underestimated the $v_2$ experimental data in the
nucleus-nucleus collisions at RHIC and/or LHC energies. In \cite{pete} it
was mentioned that the charged particle $v_2$ experimental data is around
0.05 in the Au+Au collisions at highest RHIC energy (estimated from
$v_2(\eta)$ in \cite{phob4}), while the UrQMD model provides only half of
this value. They have pointed out that a lack of pressure in the model at
this energy may be the reason and that the partonic rescattering has to
be taken into account in order to describe the data.

Similarly, the default AMPT model (AMPT$\_{def}$) also underestimated the
$v_2$ experimental data in the nucleus-nucleus collisions at RHIC energies
\cite{lin}. In order to meet with experimental data they updated
AMPT$\_{def}$ to the AMPT$\_{sm}$ with string melting. In the
AMPT$\_{sm}$ model the hadrons (strings) from HIJING \cite{wang} are
all melted to the partons. Relying on the rescattering among huge
number of partons AMPT$\_{sm}$ is able to account for the $v_2$
experimental data, provided the parton-parton cross section is enlarged
to ten mb. Of course, the AMPT$\_{sm}$ model has to hadronize the
partons after rescattering by the coalescence model rather than the
string fragmentation in AMPT$\_{def}$.

In the non-center nucleus-nucleus collisions the geometric overlap zone
leads to the initial particle spatial asymmetry distribution. It is
then dynamically developed to the final hadronic state transverse
momentum asymmetry due to the partonic rescattering \cite{pete} and
the strong electromagnetic field \cite{tuch} etc.. We have pointed out
that the transverse momentum $p_x^{'}$ and $p_y^{'}$ of produced particle
from string fragmentation are randomly arranged on a circle with radius
of $p_T^{'}$ in the PACIAE 2.0 model \cite{sa1} (in the PYTHIA model
\cite{sjos} originally). Here the observable with superscript $(')$
refers to the string fragmentation local frame distinguished from the
without superscript one referred to the nucleus-nucleus cms frame.
This symmetric arrangement strongly cancels the final hadronic state
transverse momentum asymmetry developed from the initial spatial
asymmetry. In the new issue of a PACIAE model (PACIAE 2.1 \cite{sa2})
we randomly distribute the $p_x^{'}$ and $p_y^{'}$ of produced particle
from the string fragmentation on the circumference of an ellipse instead
of circle. PACIAE 2.1 is then able to describe the $v_2$ experimental
data.

In the next section, section II, a parton and hadron cascade model
PACIAE, its new issue of PACIAE 2.1, and the definition of elliptic
flow parameter are briefly introduced. The calculated charged particle
$v_2(\eta)$ and $v_2(p_T)$ are compared with the corresponding
experimental data of the Au+Au/Pb+Pb collisions at $\sqrt{s_{NN}}$=
0.2/2.76 TeV in the section III. Additionally, the predictions for
charged particle $v_2(\eta)$ and $v_2(p_T)$ in the p+p collisions at
$\sqrt s$=7 TeV and in the p+Au/p+Pb collisions at $\sqrt{s_{NN}}$=
0.2/5.02 TeV are also given in the section III. The last section is
devoted to the conclusions.
%%%%%%%%%%%%%%%%%%%%%%%%%%%%%%%%%%%%%%%%%%%%%%%%%%%%%%%%%%%
\section {Models}
%%%%%%%%%%%%%%%%%%%%%%%%%%%%%%%%%%%%%%%%%%%%%%%%%%%%%%%%%%%
The PACIAE model is based on PYTHIA \cite{sjos}. However, the PYTHIA
model is for high energy hadron-hadron ($hh$) collisions but the PACIAE
model is mainly for nucleus-nucleus collisions. In the PYTHIA model a
$hh$ collision is decomposed into parton-parton collisions. The hard
parton-parton collision is described by the lowest leading order
perturbative QCD (LO-pQCD) parton-parton interactions with the
modification of parton distribution function in a hadron. The soft
parton-parton collision, a non-perturbative process, is considered
empirically. The initial- and final-state QCD radiations and the
multiparton interactions are also taken into account. So the consequence
of a $hh$ collision is a partonic multijet state composed of the
diquarks (anti-diquarks), quarks (antiquarks), and the gluons, besides
a few hadronic remnants. It is followed by the string construction and
fragmentation, thus a final hadronic state is obtained for a $hh$ ($pp$)
collision eventually.

In the PACIAE model \cite{sa1}, the nucleons in a nucleus-nucleus
collision are first randomly distributed in the spatial phase space
according to the Woods-Saxon distribution. The participant nucleons,
resulted from Glauber model calculation, are required to be inside the
overlap zone, formed when two colliding nuclei path through each other
at a given impact parameter. The spectator nucleons are required to be
outside the overlap zone but inside the nucleus-nucleus collision system.
Then we decompose a nucleus-nucleus collision into nucleon-nucleon ($NN$)
collisions according to nucleon straight-line trajectories and the $NN$
total cross section. Each $NN$ collision is then dealt by PYTHIA with
the string fragmentation switched-off and the diquarks (anti-diquarks)
broken into quark pairs (anti-quark pairs). A partonic initial state
(composed of the quarks, antiquarks, and the gluons) is obtained for a
nucleus-nucleus collision after all of the $NN$ collision pairs were
exhausted. This partonic initial stage is followed by a parton evolution
stage, where parton rescattering is performed by the Monte Carlo method
with $2\rightarrow2$ LO-pQCD cross sections \cite{comb}. The hadronization
stage follows the parton evolution stage. The Lund string fragmentation
model and a phenomenological coalescence model are provided for the
hadronization. However, the string fragmentation model is selected in this
calculations. Then the rescattering among produced hadrons is dealt with
the usual two body collision model \cite{sa1}. In this hadronic evolution
stage, only the rescatterings among $\pi$, $K$, $p$, $n$, $\rho (\omega)$,
$\Delta$, $\Lambda$, $\Sigma$, $\Xi$, $\Omega$, and their antiparticles
are considered for simplicity.
\begin{table*}[htbp]
\centering \caption{Charged particle pseudorapidity densities at
mid-rapidity and the fitted model parameters.}
\begin{tabular}{ccccccc}
\hline\hline
Reaction& Energy [TeV]& \multicolumn{2}{c}{$dN_{ch}/d\eta|_{mid}$}& K$^\dag$&
 $\beta^\S$& $\Delta t^\sharp$\\
\cmidrule[0.25pt](l{0.05cm}r{0.05cm}){1-1}
\cmidrule[0.25pt](l{0.05cm}r{0.05cm}){2-2}
\cmidrule[0.25pt](l{0.05cm}r{0.05cm}){3-4}
\cmidrule[0.25pt](l{0.05cm}r{0.05cm}){5-5}
\cmidrule[0.25pt](l{0.05cm}r{0.05cm}){6-6}
\cmidrule[0.25pt](l{0.05cm}r{0.05cm}){7-7}
     &     & Experiment& PACIAE&     &     & \\
\cmidrule[0.25pt](l{0.05cm}r{0.05cm}){3-3}
\cmidrule[0.25pt](l{0.05cm}r{0.05cm}){4-4}
p+p (NSD)& 0.2 & 2.25$\pm$0.33$^{1)}$& 2.08 & 1& 0.58&  \\
p+p (NSD)& 7 & 5.78$\pm$0.01$\pm$0.23$^{2)}$ & 5.74& 2& 0.58&  \\
\hline
p+Au &0.2 & & 3.63& 1& 1.7& 0.0001 \\
p+Pb (NSD)&5.02 &16.81$\pm$0.71 $^{3)}$& 16.5& 3& 0.1& 7*10$^{-4}$ \\
\hline
Au+Au &0.2 & 640$\pm50^{4)}$& 626& 1& 1.7& 0.0001 \\
Pb+Pb &2.76 & 1612$\pm$55$^{5)}$& 1659& 3& 0.1& 7*10$^{-4}$ \\
\hline\hline
\multicolumn{7}{l}{$^\dag$ Correction for the higher order and
                   non-perturbative contributions,default (D)=1.} \\
\multicolumn{7}{l}{$^\S$ A parameter in Lund string fragmentation
                   function, D=0.58.}\\
\multicolumn{7}{l}{$^\sharp$ Minimum distinguishable collision time
                   interval.}\\
\multicolumn{7}{l}{$^{1)}$ taken from \cite{phob1}, here NSD refers to the
                   non-single diffractive. \hspace{0.5cm}
                   $^{2)}$ taken from \cite{cms1}.} \\
\multicolumn{7}{l}{$^{3)}$ taken from \cite{alice}.\hspace{0.5cm}
                   $^{4)}$ taken from \cite{phob2}.} \\
\multicolumn{7}{l}{$^{5)}$ taken from \cite{cms2}.} \\
\end{tabular}
\label{mul}
\end{table*}
%\begin{widetext}
\begin{center}
\begin{figure}[]
%\vspace{2cm}
\includegraphics[width=0.3\textwidth]{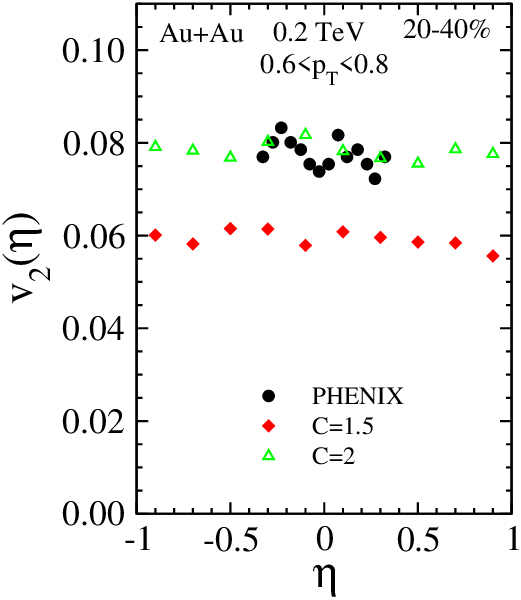} \hspace{2.5cm}
\includegraphics[width=0.3\textwidth]{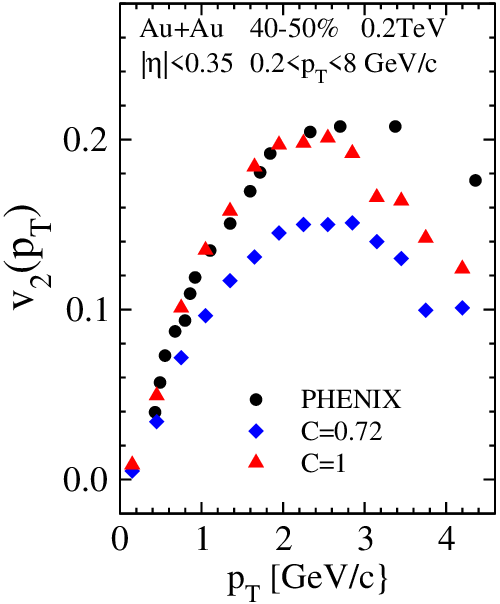}
\caption{(color on line) Charged particle $v_2(\eta)$ (left panel, 20-40\%
centrality) and $v_2(p_T)$ (right panel, 40-50\% centrality) in the Au+Au
collisions at $\sqrt s$=0.2 TeV. The PHENIX data were taken from
\cite{phen1} (using the results of event-plane method).}
\label{au_v2etapt}
\end{figure}
\end{center}
%\end{widetext}
%\begin{widetext}
\begin{center}
\begin{figure}[]
\vspace{1.5cm}
\includegraphics[width=0.3\textwidth]{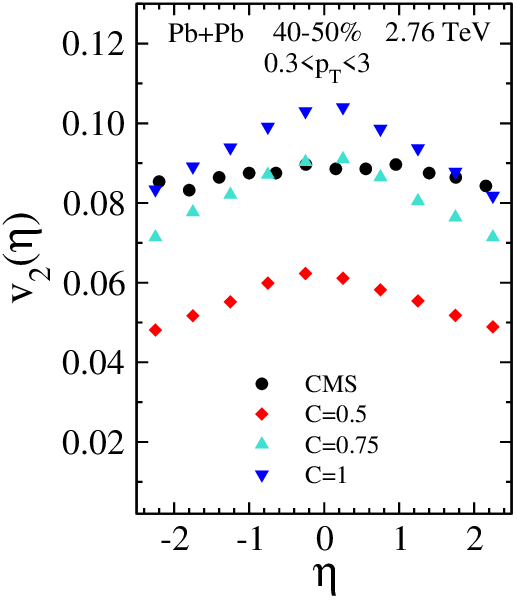} \hspace{3.0cm}
\includegraphics[width=0.3\textwidth]{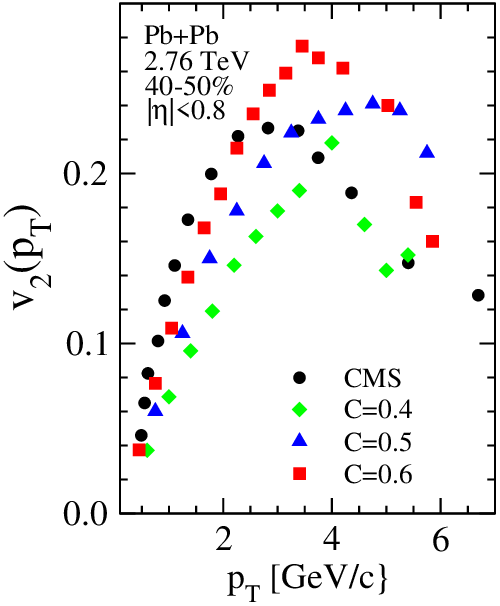}
\caption{(color on line) Charged particle $v_2(\eta)$ (left panel) and
$v_2(p_T)$ (right panel) in the Pb+Pb collisions at $\sqrt s$=2.76 TeV.
The CMS data are taken from \cite{cms} (using results of the Lee-Yang
zero point method for $v_2(\eta)$ and the event-plane method for
$v_2(p_T)$).}
\label{pb_v2etapt}
\end{figure}
\end{center}
%\end{widetext}
The PACIAE 2.0 model \cite{sa1} is mainly different from AMPT$\_{sm}$
as follows:
\begin{enumerate}
\item The partonic initial state is obtained by breaking the strings
from PYTHIA in PACIAE 2.0, but by breaking hadrons from HIJING in
AMPT$\_{sm}$.
\item The $gg\rightarrow gg$ elastic scattering cross section is
utilized in the parton rescattering in AMPT$\_{sm}$ but specific
scattering cross section is used for individual $qq$ ($gg$)
scattering processes in PACIAE 2.0 .
\item In the AMPT$\_{sm}$ model the partons after rescattering are
hadronized by the coalescent model but by string fragmentation in
the present PACIAE calculations.
\end{enumerate}
Because of the first difference, the number of initial partons in
PACIAE 2.0 is much less than the one in AMPT$\_{sm}$. Hence the
strength of partonic rescattering effect in the former is not as
strong as that in the later. Therefore relying on partonic rescattering
only the PACIAE model is hard to describe $v_2$ experimental data,
unlike AMPT$\_{sm}$. The rearrangement for the transverse momentum $x$
and $y$ components of the particles from string fragmentation, mentioned
above, is then required.

The spatial overlap zone formed in non-center nucleus-nucleus collision is
almond-like, which is always assumed to be an ellipse with a half minor
axis of $a_r=R_A(1-\delta_r)$ along the $x$ axis (axis of impact
parameter) and a half major axis of $b_r=R_A(1+\delta_r)$ along the $y$
axis (here $R_A$ refers to the radius of nucleus provided a symmetry
nucleus-nucleus collisions is considered). Originally this initial spatial
asymmetry may develop dynamically into a final hadronic state momentum
asymmetry due to the parton rescattering and the strong electromagnetic
field etc.. Unfortunately, in the PYTHIA (PACIAE 2.0) model
once the transverse momentum $p_T^{'}$ of the produced particle from
string fragmentation is randomly sampled according to the exponential
and/or Gaussian distribution, its $p_x^{'}$ and $p_y^{'}$ components are
randomly arranged on a circle with radius of $p_T^{'}$, i.e.
\begin{equation}
p_x^{'}=p_T^{'}cos(\phi^{'}),\hspace{1cm}
p_y^{'}=p_T^{'}sin(\phi^{'}),
\label{circle}
\end{equation}
where $\phi^{'}$ refers to the azimuthal angle of particle transverse
momentum. This symmetry arrangement strongly cancels the final hadronic
state transverse momentum asymmetry developed dynamically from the initial
spatial asymmetry. As a prescription to minimize this cancelation, in
PACIAE 2.1 \cite{sa2} we randomly distributed $p_x^{'}$ and $p_y^{'}$ on
the circumference of an ellipse with half major and minor axes of
$p_T^{'}(1+\delta_p)$ and $p_T^{'}(1-\delta_p)$, respectively, instead of
circle. I. e.
\begin{equation}
p_x^{'}=p_T^{'}(1+\delta_p)cos(\phi^{'}),\hspace{1cm}
p_y^{'}=p_T^{'}(1-\delta_p)sin(\phi^{'}).
\label{ellep}
\end{equation}
%\begin{widetext}
\begin{center}
\begin{figure}[]
%\vspace{1cm}
\includegraphics[width=0.6\textwidth]{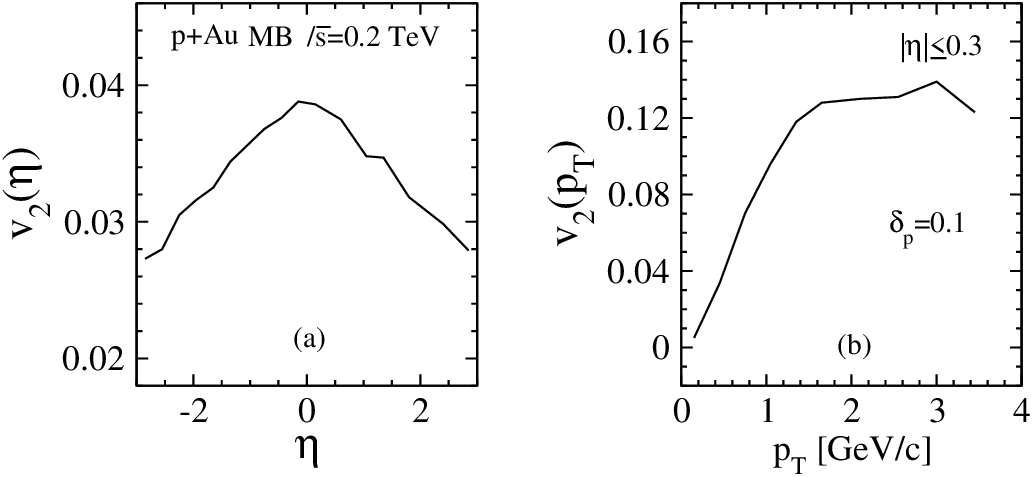}
\caption{Predicted charge particle $v_2(\eta)$ (panel (a)) and $v_2(p_T)$
((b)) in the p+Au collisions at $\sqrt{s_{NN}}$=0.2 TeV ($\delta_p$=0.1).}
\label{pau_v2etapt}
\end{figure}
\end{center}
%\end{widetext}

We know from ideal hydrodynamic calculation \cite{kolb} that the
integrated elliptic flow parameter of final hadronic state is
approximately proportional to the initial spatial eccentricity of nuclear
overlap zone. Therefore we assume that the introduced deformation
parameter of $\delta_p$ here can be related to the deformation parameter
of $\delta_r$ in the initial spatial phase space, i. e.
\begin{equation}
\delta_p=C\delta_r
\label{delt}
\end{equation}
where $C$ is an extra model parameter instead of $\delta_p$. We also
know that the spatial eccentricity of nucleon distribution in the
initial overlap zone, reaction plane eccentricity for instance
\cite{phob3}, can be expressed as
\begin{equation*}
\epsilon_{rp}=\frac{\sigma_y^2-\sigma_x^2}{\sigma_y^2+\sigma_x^2},
\end{equation*}
\begin{equation*}
\sigma_x^2=<x^2>-<x>^2,
\end{equation*}
\begin{equation}
\sigma_y^2=<y^2>-<y>^2,
\label{eccc}
\end{equation}
where $<...>$ denotes the average over the nucleon spatial distribution.
This spatial eccentricity should be identical with the geometrical
eccentricity \cite{beye}
\begin{equation}
\epsilon_g=\sqrt{\frac{b_r^2-a_r^2}{b_r^2}}
\label{eccg}
\end{equation}
of the ellipse of initial spatial overlap zone. Using $\epsilon_{rp}$
instead of $\epsilon_g$ on the left hand side of Eq.~(\ref{eccg}) and
inserting $b_r=R_A(1+\delta_r)$ as well as $a_r=R_A(1-\delta_r)$ on the
right hand side of Eq.~(\ref{eccg}), we obtain an algebraic equation of
degree 2 in the unknown $\delta_r$
\begin{equation}
\epsilon_{rp}^2 \delta_r^2+(2\epsilon_{rp}^2-4)\delta_r+\epsilon_{rp}^2
 =0.
\label{algeq}
\end{equation}
This equation has two analytical roots: The one less than unity is a
physical root
\begin{equation}
\delta_r=\frac{2-\epsilon_{rp}^2-2\sqrt{1-\epsilon_{rp}^2}}
              {\epsilon_{rp}^2}.
\label{delt2}
\end{equation}
Another one larger than unity is an unphysical root because $\delta_r$
must be $\leq 1$. The approximation of
\begin{equation}
\delta_r\simeq\frac{\epsilon_{rp}^2}{4}
\label{delt1}
\end{equation}
introduced in PACIAE 2.1 \cite{sa2} is just a specifically approximated
root of Eq.~(\ref{algeq}). For the p+p and p+A collisions, the weak initial
spatial fluctuation (asymmetry) is also possible to be dynamically
developed to the final hadronic state transverse momentum asymmetry
and the Eq.~(\ref{delt}) steamed from hydrodynamic calculation \cite{kolb}
may also be reliable. Just because of the lack of a proper definition for
the initial spatial fluctuation (eccentricity ?), we regard $\delta_p$
itself as an extra model parameter temporarily.
%\begin{widetext}
\begin{center}
\begin{figure}[]
\vspace{1cm}
\includegraphics[width=0.6\textwidth]{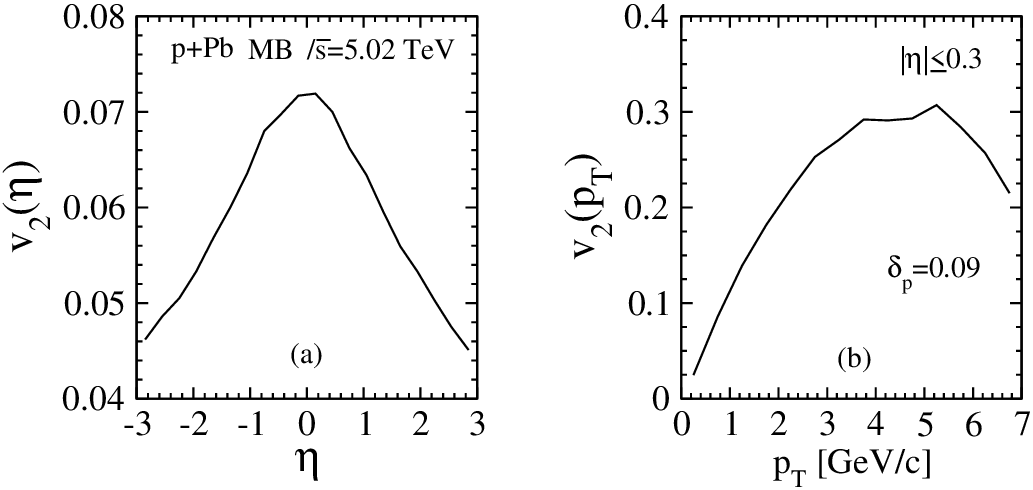}
\caption{Predicted charge particle $v_2(\eta)$ (panel (a)) and $v_2(p_T)$
((b)) in the p+Pb collisions at $\sqrt{s_{NN}}$=5.02 TeV ($\delta_p$=0.09)
.}
\label{ppb_v2etapt}
\end{figure}
\end{center}
%\end{widetext}

The Fourier expansion of particle transverse momentum azimuthal
distribution reads
\cite{posk,zhang}
\begin{equation}
E{\frac{d^3N}{d^3p}}=\frac{1}{2\pi}\frac{d^2N}{p_Tdydp_T}
       [1+\sum_{n=1,2,...}2v_ncos[n(\phi-\Psi_r)],
\label{f1}
\end{equation}
where $\phi$ refers to the azimuthal angle of particle transverse
momentum, $\Psi_r$ stands for the azimuthal angle of reaction plane. In
the theoretical study, if the beam direction and impact parameter vector
are fixed, respectively, on the $p_z$ and $p_x$ axes in the
nucleus-nucleus cms frame, then the reaction plane is just the $p_x-p_z$
plane \cite{zhang}. Therefore the reaction plane angle, $\Psi_r$, between
the reaction plane and the $p_x$ axis \cite{zhang} introduced for
extracting the elliptic flow experimentally \cite{posk} is zero. The
equation (\ref{f1}) and the harmonic coefficients there reduce to
\begin{align}
E{\frac{d^3N}{d^3p}}=&\frac{1}{2\pi}\frac{d^2N}{p_Tdydp_T}
       [1+\sum_{n=1,2,...}2v_ncos(n\phi)], \nonumber\\
\langle v_n\rangle_p=&\langle cos(n\phi)\rangle_p, \nonumber\\
\langle v_1\rangle_p=&\langle\frac{p_x}{p_T}\rangle_p, \nonumber\\
\langle v_2\rangle_p=&\langle\frac{p_x^2-p_y^2}{p_T^2}\rangle_p,
 \nonumber\\
 . . .
\label{f2}
\end{align}
where $\langle ... \rangle_p$ denotes the particle-wise average, i.e.
the average over all particles in all events \cite{posk}.
%%%%%%%%%%%%%%%%%%%
\section {Results and discussions}
%%%%%%%%%%%%%%%%%%%
In the PACIAE 2.1 model simulations, the model parameters are all fixed
as the same as default values given in PYTHIA, except the $K$ factor,
$\beta$, and $\Delta t$. They are, respectively, the higher order term
corrections for the LO-pQCD parton-parton cross section \cite{sjos}, a
factor in the Lund string fragmentation function \cite{sjos}, and the
least time interval of two distinguishably consecutive collisions in the
partonic initial and evolution stages \cite{sa1}. These model parameters
are first fitted to the experimental data of charged particle
pseudorapidity density and are given in Tab.~\ref{mul}. Later on, these
fitted parameters are used in all of the simulations. Additionally, in
this study the participant eccentricity \cite{phob3} of
\begin{equation}
\epsilon_{pa}=\frac{\sqrt{(\sigma_y^2-\sigma_x^2)^2+4\sigma_{xy}^2}}
                   {\sigma_y^2+\sigma_x^2}
\label{ecccp}
\end{equation}
is used instead of reaction plane eccentricity $\epsilon_{rp}$. In the
above equation $\sigma_{xy}$ is equal to $<xy>-<x><y>$. Meanwhile, the
physical root of Eq. (\ref{delt2}) is employed instead of the specifically
approximated root of Eq. (\ref{delt1}).
%\begin{widetext}
\begin{center}
\begin{figure}[]
\includegraphics[width=0.6\textwidth]{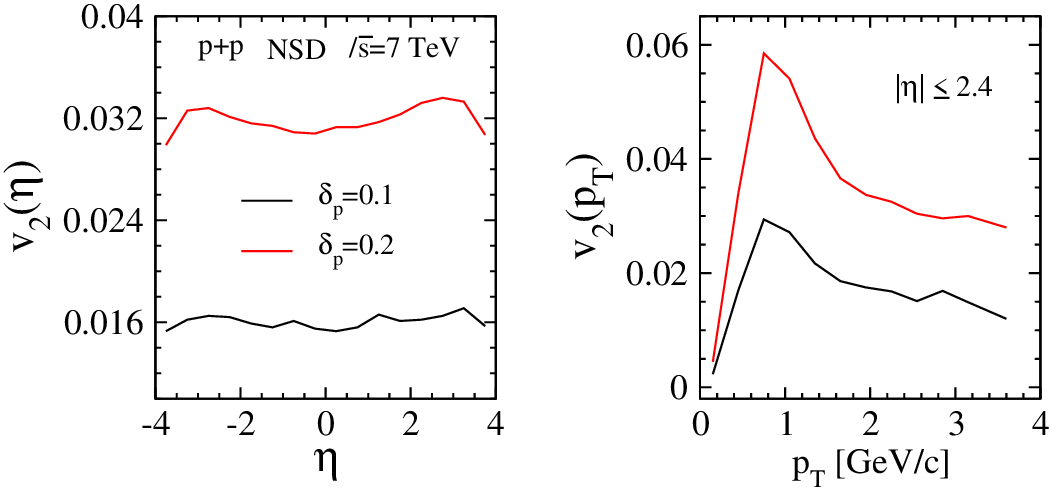}
\caption{(color on line) Predicted charge particle $v_2(\eta)$ (left panel)
and $v_2(p_T)$ (right) in the p+p collisions at $\sqrt s$=7 TeV.}
\label{pp7_v2etapt}
\end{figure}
\end{center}
%\end{widetext}

We compare the calculated charged particle $v_2(\eta)$ and $v_2(p_T)$ in
the 20-40\% and 40-50\% central Au+Au collisions at $\sqrt{s_{NN}}$=0.2
TeV with the corresponding experimental data in the left and right panels
of Fig. \ref{au_v2etapt}, respectively. The PHENIX data were taken from
\cite{phen1} (using the results of event-plane method). One sees in the
left panel that the PACIAE 2.1 results calculated by $C$=2 well agree
with the PHENIX data. The right panel shows that the model results
calculated by $C$=1 reproduce PHENIX data quite well in the $p_T<$ 3
GeV/c region. However, the theoretical result decreases with $p_T$
increasing is faster than experimental data in the $p_T>$3 GeV/c region.
As most of particles are generated below $p_T\sim$2 TeV/c (about 95
percent of the total multiplicity), one always satisfies the
agreement between model calculations and experimental data within
$p_T\leq$2 GeV/c, cf. Fig. 7 in the first quotation of Ref.~\cite{lin}
for instance. As for the best model parameter $C\sim$2 in the left
panel but 1 in the right panel, which may be attributed to the
difference in the studied centrality bin, 20-40\% in former but
40-50\% in the later. Thus the centrality dependence of parameter $C$
should be studied later.

Similarly, the calculated charged particle $v_2(\eta)$ and $v_2(p_T)$ in
the 40-50\% central Pb+Pb collisions at $\sqrt{s_{NN}}$=2.76 TeV are
compared with the corresponding CMS data \cite{cms} (using the results
of Lee-Yang zero point method for $v_2(\eta)$ and event-plane method for
$v_2(p_T)$) in Fig. \ref{pb_v2etapt}. We see in this figure that the
PACIAE 2.1 model is also able to describe the CMS data by adjusting
the extra parameter $C$.

In the Figures \ref{pau_v2etapt}, \ref{ppb_v2etapt}, and \ref{pp7_v2etapt}
we give the PACIAE 2.1 model predictions for the charged particle
$v_2(\eta)$ and $v_2(p_T)$ in the minimum bias (MB) p+Au and p+Pb, as
well as in the non-single diffractive (NSD) p+p collisions at
$\sqrt{s_{NN}}$=0.2, 5.02, and 7 TeV, respectively. We see in
these figures that the elliptic flow parameter may reach a amount of
0.04, 0.07, and 0.016 (estimated from $v_2(\eta)$) in the p+Au, p+Pb,
and p+p collisions at $\sqrt{s_{NN}}$=0.2, 5.02, and 7 TeV, respectively.
This amount of the elliptic flow parameter may be measurable experimentally
. One sees in Fig. \ref{pp7_v2etapt} that $v_2$ seems to be proportional
to the value of deformation parameter $\delta_p$ in the p+p collisions.
However, the behavior of $v_2(\eta)$ and $v_2(p_T)$ changing with
$\delta_p$ is needed to be further investigated in detail.
%%%%%%%%%%%%%%%%%%%%%%%%%%%%%%%%%%%%%%
\section {Conclusions}
%%%%%%%%%%%%%%%%%%%%%%%%%%%%%%%%%%%%%%
In summary, We have employed the new issue of a parton and hadron
cascade model PACIAE 2.1 investigating systematically the charged
particle elliptic flow parameter $v_2$ in the relativistic nuclear
collisions at RHIC and LHC energies. Because of the new introduced
mechanism of random arrangement of the particles from string
fragmentation on the circumference of an ellipse instead of circle
originally, the calculated charge particle $v_2(\eta)$ and $v_2(p_T)$
in the Au+Au/Pb+Pb collisions at $\sqrt{s_{NN}}$=0.2/2.76 TeV describe
the corresponding experimental data fairly well. Meanwhile, the
charged particle $v_2(\eta)$ and $v_2(p_T)$ in the p+Au/p+Pb collisions
at $\sqrt{s_{NN}}$=0.2/5.02 TeV and in the p+p collisions at
$\sqrt s$=7 TeV are predicted. The elliptic flow parameter
in these reactions reaches a measurable amount.

As mentioned in the first section that the elliptic flow parameter
is important observable relevant to the exploring of sQGP. However,
the measurement of $v_2$ is not trivial. The discrepancy among $v_2$
values measured by the event plane method \cite{posk}, Lee-Yang zero point
method \cite{lyzp}, and the cumulant method \cite{cumu} may reach a few
ten percent as shown in Fig. 4 and 5 of \cite{star1} and Fig. 11 of
\cite{cms}. On the other hand, the obscures also exist among the various
$v_2$ model calculations as mentioned in the first section. So the further
studies for $v_2$ asymmetry etc. are still required both experimentally
and theoretically.

This work is just a first step along the novel approach. Further
investigations, such as the cross section effect, energy and centrality
dependence of $C$ parameter, as well as the detail study for the
dependence of $v_2(p_T)$ and $v_2(\eta)$ on $\delta_p$ ($C$) etc.,
are really required.

Acknowledgements: This work was supported by the National Natural Science
Foundation of China under grant Nos.:11075217, 11105227, 11175070,
11477130 and by the 111 project of the foreign expert bureau of China.
BHS would like to thank Zi-Wei Lin for discussions. YLY acknowledges the
financial support from SUT-NRU project under contract No. 17/2555.
%\newpage

\end{document}